# High-Speed Light Focusing through Scattering Medium by Cooperatively Accelerated Genetic Algorithm

SHU GUO,[1,2] AND LIN PANG[1,2]

[1]College of Physics, Sichuan University, No. 24 South Section 1, Yihuan Road, Chengdu, 610065, People's Republic of China
[2] LinOptx LLC, 6195 Cornerstone Court, San Diego, CA 92121, USA
*pang@linoptx.com

**Abstract:** We develop an accelerated Genetic Algorithm (GA) system constructed by the cooperation of field-programmable gate array (FPGA) and optimized parameters of the GA. We found the enhanced decay of mutation rate makes convergence of the GA much faster, enabling the parameter-induced acceleration of the GA. Furthermore, the accelerated configuration of the GA is programmed in FPGA to boost processing speed at the hardware level without external computation devices. This system has ability to focus light through scattering medium within 4 seconds with robust noise resistance and stable repetition performance, which could be further reduced to millisecond level with advanced board configuration. This study solves the long-term limitation of the GA, it promotes the applications of the GA in dynamic scattering mediums, with the capability to tackle wavefront shaping in biological material.

## 1. Introduction

Imaging through scattering mediums has long been a significant challenge in decades. It attracts lots of research attention since it offers advantages for biomedical diagnosis and therapy. However, the light transmitted through the scattering medium (like biological tissues) experiences multiple scattering processes, which leads to limited superficial depth. During the scattering process, the ballistic and quasi-ballistic photons (single scattered photons) that carry image information would degrade exponentially. To manipulate these photons and enable imaging in biological tissues, conventional microscopy is employed to gather ballistic photons when propagation distance is smaller than the scattering mean free path, which accomplishes the diffraction-limit imaging. But as the propagation distance or depth increases, the transmitted field would be further randomized, in which the scrambled speckles are formed behind. In this regime, the intensity of ballistic photons would be too weak to be detected. To tackle the highly scattering issue, researchers have attempted to manipulate scattered photons to interfere constructively. In the past years, emerged wavefront shaping technologies have been developed to control incident wavefront to form light focusing behind the scattering medium. Three common approaches of wavefront shaping are transmission matrix (TM) [1-5], optical phase conjugation (OPC) [6-9], and feedback-based iterative wavefront shaping [10]. The TM represents properties of the scattering medium, in which the conjugation of one column of TM could optimize the scattered field at one channel. However, the efficiency of TM is highly reliant on the signal-to-noise ratio (SNR) of measurement systems [11], and the evaluation process requires the scattering medium to maintain static. The OPC directly measures the scattered optical field, reverses its phase to make scattered light propagate back. Although OPC could restore incident optical field in milliseconds, the 'guide star' and highly stable optical system are needed. The iterative wavefront shaping is accomplished by computational algorithms that gradually optimize the incident optical field according to the measured light intensity values as feedbacks. For the purpose of light focusing, various algorithms for iterative wavefront shaping were studied, such as partitioning algorithm (PA) and continuous sequential algorithm (CSA) [12], Hadamard algorithm [13], simulated annealing algorithm [14], particle swarm algorithm (PSA) [15,16], four-element division [17] and genetic algorithm (GA) [11,18,19]. However, with different cons and pros as mentioned before, wavefront shaping approaches have not been widely applied in dynamic scattering regimes, especially for in-vivo light focusing applications. These applications require properties including fast (match the scattering decorrelation time), stable (could be repetitively used), and noise robustness. The OPC and TM approaches might offer fast retrieval of optimized field, but their evaluation principles determine their significant weakness against noise, making them unsuitable for dynamic scattering. Among wavefront shaping approaches, the GA could achieve higher enhancement of light focusing in weak SNR conditions due to its advanced property of being immune to the noisy environment compared to other methods. But the GA is a random search-based optimization algorithm that achieves the global optimum after multiple iterative operators, including selections, crossovers, and mutations, etc. The computational complexity of the GA makes it highly efficient for optimization, but at the cost of consuming much longer time. As a result, the GA implemented on a personal computer requires typically at least dozens of minutes (or even several hours). The time-consuming issue makes GA an unusable strategy for applications in dynamic scattering medium currently. Consequently, if the timing issue of the GA can be resolved, the fast, stable, and noise robustness requirements of applications in dynamic scattering would be satisfied, in which the GA could be readily used. To achieve this goal, decreasing the total iteration number, or speeding up convergence of the GA, while reducing processing time for each iteration, would be specific strategies that must be tackled.

Studies have shown that selections of the GA parameters could affect performance of the GA [19,20]. Within the GA parameters, the mutation rate is significant as it determines the size of search space, and further influences convergence of the GA. For example, the initial mutation rate shows an inversely proportional relation with convergence speed. Unfortunately, a low initial mutation rate might result in 'genetic drift' and make the GA challenging to achieve the global optimum in a noisy environment. To enable the faster and stabler convergence of the GA, we propose to adopt a sufficient initial mutation rate and a higher decay ratio of mutation rate. The higher decay ratio shrinks search space by declining mutation rate and shortens convergence time of the GA. On the other hand, we significantly accelerate the GA operation in the hardware base. Wavefront shaping technologies are normally achieved by employing a spatial light modulator (SLM), including liquid crystal based SLM (LC-SLM) and digital micromirror device (DMD). The LC-SLM is more commonly used since it can continuously modulate optical phase, but LC-SLM`s frame rate is typically only 60Hz. By contrast, DMD could reach up to 22kHz but with binary amplitude modulation, although it might lose the overall modulation efficiency. For the consideration of timing reduction, DMD would be adopted as a high-speed SLM in this study. More importantly, the operators of GA like selection, crossover, and mutation would be implemented on the field-programmable gate array (FPGA) chip. FPGA is a hardware circuit that could be programmed for multiple logic functions. In wavefront shaping applications, previous studies have reported using FPGA to form focal spots by closed-loop algorithm [21] and TM approach [22]. According to our previous designs, the FPGA-based TM approach with $32 \times 32$ modulation units could generate focal spots within around 132ms. However, the TM approach is exceptionally susceptible to environmental noise, let along higher enchantment of focal spot will cost more time as it acquires higher-dimensional TM with more modulation units. Consequently, in this paper, we aim to design an FPGA program that accomplishes the control of DMD and GA operation concurrently, which diminishes the processing time of each iteration of the GA in hardware base.

We first analyze the effect of decay ratio to convergence time of the GA by conducting a series of numerical simulations as well as experimental evaluations, through which we define the optimized parameters. These parameters were then adopted into FPGA design to build

FPGA-based GA configuration, including DMD data transferring and displaying, write/read operation of Double Data Rate Synchronous Dynamic Random-Access Memory (DDR SDRAM)), acquisition of digitalized data from an analog-to-digital converter (ADC). Under the collaboration between FPGA and optimized parameters, a focal spot is generated by the accelerated GA within a few seconds with decreased number of iterations and reduced processing time for each iteration, which speeds up about 150 times than computer-based systems. With further upgrade of hardware, we believe the proposed accelerated GA approach could reach the level of milliseconds, suitable for most dynamic scattering media.

## 2. Rapid Convergence of Genetic Algorithm with Optimized Parameters

### 2.1 Theory behind Convergence of GA

The GA is an optimization algorithm that searches for optimal solutions based on nature of evolution. GA could be used to focus the scattered light to one output channel behind scattering medium by optimizing the input modulation modes to construct in-phase interference. The overall processing procedures of the GA were described in [18]. In general, the GA starts by generating an initial population of parent masks with random field (amplitude or phase), each mask also stands for a solution of optimization. Displaying the population of parent masks on the scattering medium, using a cost function to evaluate fitness of each mask. The cost function here is the ranking of intensity values measured from a detector behind scattering medium. According to the ranking of current population, two higher-ranked parent masks are selected to breed a new offspring mask, this process is called crossover. The new offspring mask is mutated by randomly changing values of a set number of input modes. New offspring masks are continuously generated from higher-ranked parent masks and projected onto scattering medium, and their corresponding intensity values are ranked again. The best-ranked half of offspring masks would replace the worst half parent masks to form a new population, the algorithm then iterates for next measurement and continues to search global optimum, the global optimum here refers to the mask that generates highest enhancement of focal spot.

Our goal is to greatly reduce the convergence time while maintaining the high level of enhancement factor (overall efficiency) of focal spot. For this purpose, the mutation rate $R$, as the ratio of random change of the modes in each mask, is a critical parameter. At each iteration of GA, mutation rate decides the diversity of the newly generated population, or defines the size of search space. When mutation rate $R$ is high, the size of search space is increased, which would help the algorithm jump out those possible local optimums. This also makes the searching of the global optimum becomes difficult. Inversely, if $R$ was relatively low, it would make the searching process faster, but might lead to 'genetic drift'. As a result, there is a trade-off between high and low mutation rates. To accelerate the searching of global optimum while avoiding being trapped in local optimums, we consider the decay ratio of mutation rate for each iteration, which is defined as:

$$\Upsilon = \{R_k - R_{k+1}\} \tag{1}$$

The decay ratio is determined by the difference of mutation rates between the $k^{th}$ and $(k+1)^{th}$ iteration. The higher decay ratio and reasonable initial mutation rate could lead to rapid convergence in GA by quickly reducing the size of search space while preventing being trapped in any potential local optimums. As shown in Fig. 1, the global optimum searching process of GA is visualized with a 2D Shaffer function, which includes many local optimums and local minimums, and one global optimum at central circle. At the beginning of GA, the mutation rate would be a bit larger, for example, 5%~15%, in order to maintain a relatively larger search space, in which higher diversity of masks could be found in initial population (larger area occupied by initial population in Fig. (1). When the GA continues to iterate, the mutation rates keep decreasing with a high decay ratio $\Upsilon$, in which the search space is shrinking rapidly like red circles show, so that the convergence speed of the GA is faster. In order to evaluate the above theory, we conduct numerical simulations and experimental assessments.

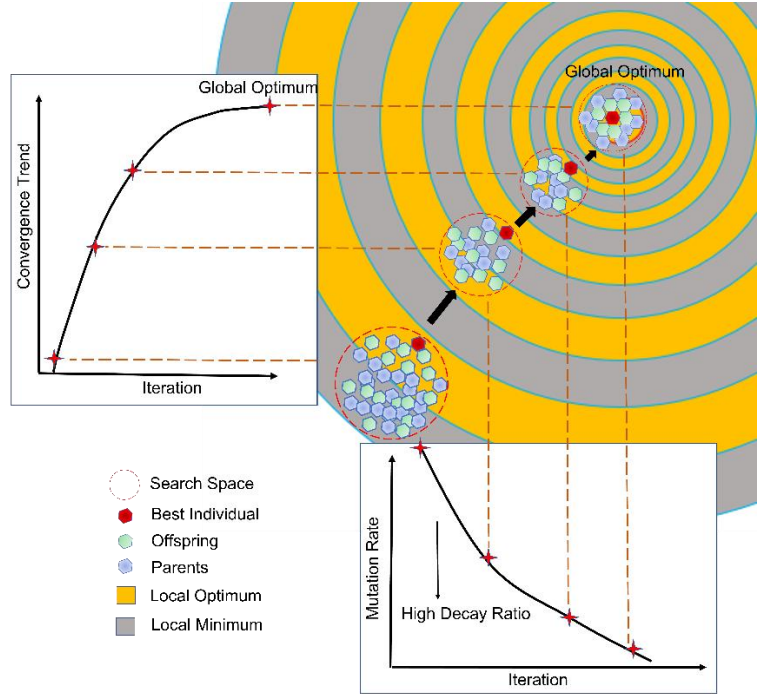

Fig.1 Description of effect of high decay ratio. The 2D Schaffer function that is used for testing black-box optimization algorithms is plotted as multiple concentric circles, the orange area indicates local optimums, the gray area represents local minimums, the global optimum is located at the central point of circle. For the GA, the diversity of a population determines how vast the search space could be, which is shown as the area of red circle. Green hexagons and blue hexagons represent offspring and parent masks, respectively. With high decay ratio, mutation rate decreases quickly along iterations and reduces the size of search space. This process improves the possibility to reach global optimum without falling into local optimums, which thus makes the convergence of GA faster.

*2.2 Numerical Simulation*

The numerical simulation is conducted in MATLAB (2019b) on a computer equipped with Window 10 OS, Intel (R) Core (TM) i9-9900K CPU @ 3.60GHz, 3600MHz, 8 Core, and 64.0 GB RAM. The number of input modulation modes is set as 1024, corresponding to a mask with $32\times32$ pixels, on which the binary amplitude modulation of DMD is simulated. The scattering medium is defined by a Gaussian random matrix. The size of one population in the GA is set as 32, meaning there are 32 masks. The cost function is the ranking of intensity values at the targeted output channel behind scattering medium, corresponding to the fitness of 32 masks. According to a previous study [18], the mutation rate $R$ for each iteration is defined as:

$$R = (R_0 - R_{end}) \times \exp(-k/D) + R_{end} \qquad (2)$$

Where $R_0$ is initial mutation rate, $R_{end}$ is ending mutation rate. $D$ is the decay factor that decides the decay ratio of mutation rate. The decay factor is inversely proportional to decay ratio, in which $\Upsilon \propto (1/D)$. To compare the effect of decay ratio, the decay factor is set as $D = 80, 400, 1000$, respectively, to indicate high, medium, and low decay ratio. The other parameters settings are given as follows: initial mutation rate $R_0 = 0.06$; ending mutation rate $R_{end} = 0.012$; total iteration number $N = 2000$. The performance of the GA with different decay ratios is evaluated by comparing normalized convergence. The normalized convergence is described as the percentage of the enhancement when global optimum is approached the highest enhancement in each iteration. Here, the enhancement $\xi$. is defined as the ratio of spot intensity after optimization to the averaged speckles` intensity before optimization. Global optimum is determined by the highest enhancement from simulated results with three decay ratios. Fig. 2 (a) illustrates the mutation rate for high, medium, and low decay ratio. For high decay ratio, the mutation rate decreases drastically to the ending mutation rate near the 500$^{th}$ iteration. In contrast, the mutation rate for medium decay ratio reaches ending mutation rate at around the 2000$^{th}$ iteration. The mutation rate for low decay ratio is far above ending mutation rate at the 2000$^{th}$ iteration. Fig. 2 (b) shows the performance of GA with mutation rates in Fig. 2 (a). The right axis indicates the highest enhancement of focal spot in each iteration, while the left axis shows normalized convergence. The global optimum appears at around the 1600$^{th}$ iteration for high decay ratio with enhancement of 96. It can be seen from Fig. 2 (b), the improved speeds of normalized convergences are high at the beginning for three decay ratios like the black dotted lines depict, and the normalized convergences reach around 20% within only about 50 iterations. The first 50 iterations refer to the first stage of improved speeds for medium and low decay ratios. After that, the improved speeds show different trends for three cases. For high decay ratio, the improved speed maintains till the normalized convergence reaches 66% at the 200$^{th}$ iteration, which corresponds to the first stage of the improved speed. Starting from the 200$^{th}$ iteration, the improved speed slows down until the 500$^{th}$ iteration with 88% normalized convergence, as the second stage. After the 500$^{th}$ iteration, the improved speed further decreases till reaching saturation at around the 2000$^{th}$ iteration as the third stage. As a contrast, improved speed of normalized convergence for medium and low decay ratio starts to decrease during their second stages, the normalized convergence reaches 70% at about 700$^{th}$ iteration for medium decay ratio, while normalized convergence for low decay ratio arrives 50% at the 570$^{th}$ iteration. Later, during their third stages, the normalized convergences finally reach 98% and 86%, for medium decay ratio and low decay ratio, respectively, corresponding to enhancement 95 and 81. The simulation results in Fig. 2 show that, in case of scattering focusing, steeper decay of mutation rate will lead the GA to converge faster and shorten the convergence time. As illustrated in Fig.1, faster convergence of the GA results from the rapid shrink of search space. Although decay ratio, or decay factor, as mentioned in reported literature, its effect on the convergence of the GA is first presented in detail above.

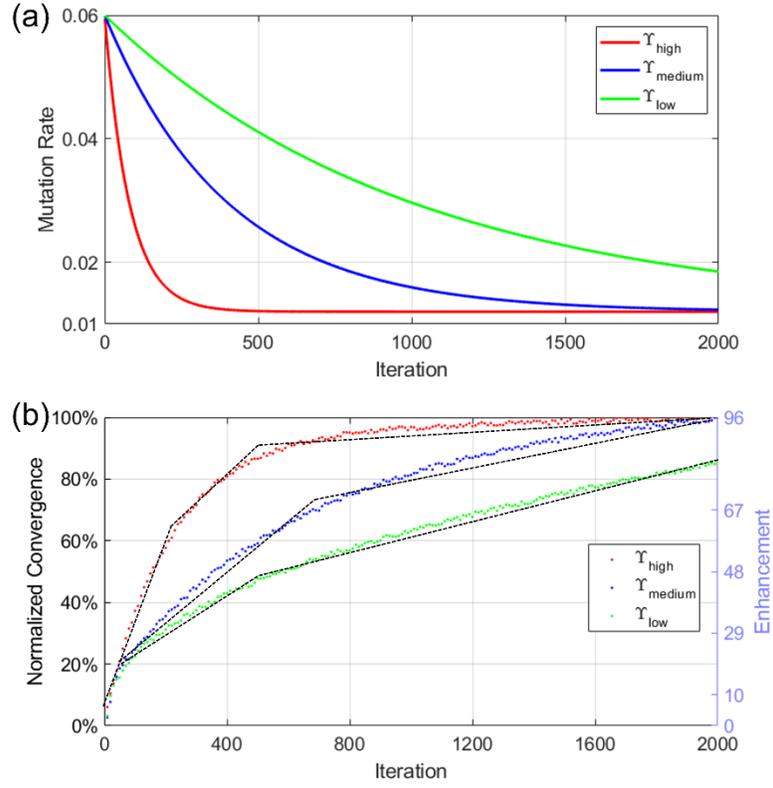

Fig.2 Numerical simulation of GA with different decay ratio. (a) Mutation rates in each iteration of the GA with decay factors equal to 80, 400, and 1000, corresponding to high, medium, and low decay ratio. (b) The normalized convergence of GA and corresponding enhancement. The three stages of convergence rates are depicted by black dotted lines.

*2.2 Experiment Validation for Effect of Optimized Parameter*

To experimentally demonstrate the evaluation of effect of decay ratio, we conducted experiments. The setup is illustrated in Fig. 3, in which a 532nm coherent laser beam (Genesis MX 532, Coherent Inc.) is expanded and illuminates on a DMD (DLP® Discovery 4100 with DLPLCRC Modulation Board DLPLCRC410EVM and DMD board, DLPLCR70EVM, Texas Instruments). The incident angle on the DMD surface is adjusted to make the micromirrors act as a blazed grating to enable the optical energy to concentrate on the reflection direction [23]. The surface of DMD is imaged on scattering medium (ground glass diffuser, DG10-120, Thorlabs) by a 4-f imaging system with a demagnification of 4 determined by focal lengths of 300mm and 75mm for the lens L2 and L3, respectively. A 90:10 beam splitter (BS) is placed behind scattering medium, from which the transmitted light, or 90% of the scattered light, is received by a photodetector (APD130A2, Thorlabs) with a $100\mu m$ iris for intensity evaluation.

The reflected light from 10% of scattered light, as a sampled light, is captured by a CCD (CS2100M-USB, Thorlabs) to visualize the formed focal spot. The analog value of voltage from the detector is transferred to a data transition box (DAQ, DT9834) which has a data resolution of 16-bit in the range of 20V from -10V ~ +10V. The DAQ is connected to a PC by USB, the same computing device is used for numerical simulation in above section. Besides,

DMD is also connected to PC and controlled by a MATLAB control module (LO4100, LinOptx LLC).

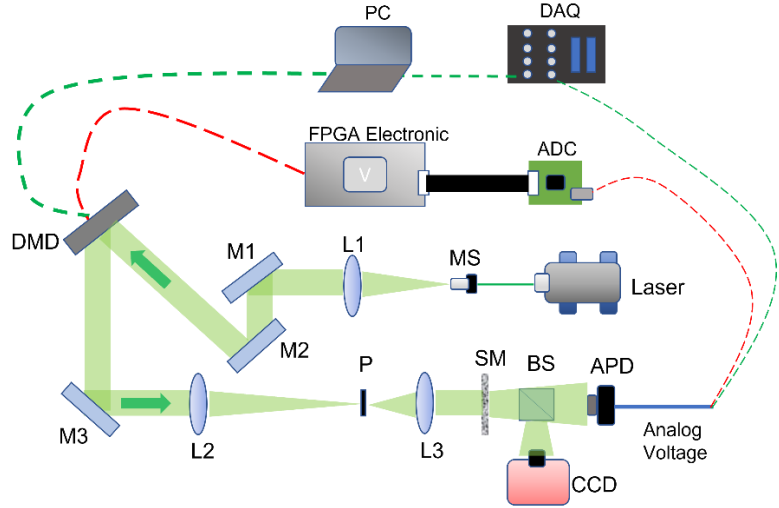

Fig. 3 The experiment setup. DMD: digital mirror device (DLP4100, TI). M1, M2, M3: mirror. L1: lens (focal length = 300mm). L2: lens (focal length = 300mm). L3: lens (focal length = 75mm). MS: microscope (M-5X, Newport). P: pinhole. SM: ground glass diffuser (DG10-120, Thorlabs). BS: beam splitter 90:10. CCD: CCD camera (CS2100M, Thorlabs). APD: Si avalanche photodetector (APD130A2, Thorlabs). DAQ: data transition box (DT9834). ADC: analog to digital convert (LinOptx Digitizer v1.0). FPGA: Virtex-5 (Xilinx). The coherent 532nm laser beam is expanded by a microscope and collimated by L1. The DMD is imaged to front surface of scattering medium by a 4-f system, the scattered light behind scattering medium is collected by a photodetector and a camera. The voltage signal from photodetector is split, one is connected to the PC control module (DAQ and PC) via green dotted line, the other one is directed to the FPGA control module (ADC and FPGA) via red dotted line. The experiment setup has two functions, the first is to validate effect of decay ratio in an experimental environment. The second one is to confirm the design of FPGA program and visualize performance.

As an experimental proof of concept, the parameters are the same as those in the simulations. Decay factors are selected as 80, 400, and 1000 to indicate high, medium, and low decay ratios. The DMD full-screen pixels in size of $1024 \times 768$ are grouped into big segments to create $32 \times 32$ input modes as one mask. Laser power is set to 45mW. Fig. 4 (a) shows the mutation rates at each GA iteration, which is the same as in Fig. 2 (a). The normalized convergence of the GA and focal spot enhancement in experiments for different decay ratios are shown in Fig. 4 (b). Basically, the experimental results follow the simulated trends shown in Fig. 2 (b) except for slight differences. In simulations, after a rapid increase of normalized convergence in the first stage from the 1$^{st}$ to 50$^{th}$ iteration, the improved speed of convergence for high decay ratio keeps relatively high level. The medium and low decay ratio regard to declination of improved speed. In experiments, the first stage of improved speed becomes short for cases of medium decay ratio and low decay ratio, is about only 30 iterations, the normalized convergence reaches 40% for three decay ratios. Afterward, the improved speed starts to separate. For high decay ratio, the improved speed decreases at the iteration of 230 till the 500$^{th}$ iteration, where the normalized convergence reaches 90%. Later, the improved speed further falls, and the GA converges to global optimum at the 2000$^{th}$ iteration with the enhancement of 43. As for medium decay ratio, the improved speed decreases from the 30$^{th}$ iteration to 580$^{th}$ iteration, which is realized as the second stage. It reaches 83% of normalized convergence at the 580$^{th}$ iteration.

The improved speed is much lower while the GA converges to the enhancement of 41, it approaches about 95% of the normalized convergence during the third stage. Referred to low decay ratio, the improved speed decreases during the 30$^{th}$ iteration to 410$^{th}$ iteration while normalized convergence reaching 64%, as the second stage. After the 410$^{th}$ iteration, the improved speed further declines and leads to 90% convergence of GA at the end. The convergence of GA might not reach global optimum, as 100% normalized convergence, for the cases with a low value of decay ratio, even though it cost a vast number of iterations. The relatively higher mutation rate would be responsible for it, because it is more difficult to obtain optimized masks under big search space.

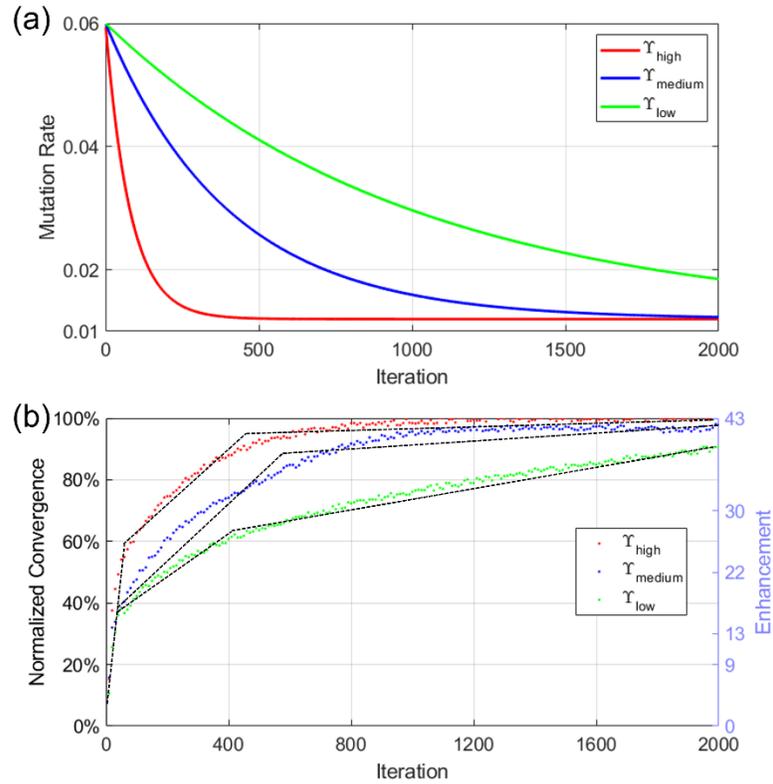

Fig. 4 Experiment validation for three decay ratios. The parameters of GA are same as numerical simulation. (a) Mutation rate of the GA experiment. (b) Normalized convergence and corresponding enhancement during measurement. The black dotted lines depict three stages of improved speed for each decay ratio setting.

## 2.3 Consuming Time versus Operational Enhancement

As illustrated in Fig. 3 and Fig. 4, the GA converges to the global optimum in a specific number of iterations. With different selections of parameters, the improved speeds of convergence indeed decrease through three stages, while the convergence of the GA gradually slows down. This brings a sacrifice of timing efficiency because the GA achieves a small quantity of enhancement improvement but consumes much longer time. In the applications of dynamic scattering, the short decorrelation time of scattering events does not allow the GA to process too many iterations. The specified timing should be determined to end the optimization processing. Upon here, the optimized mask would be displayed on DMD to form light focusing. Hence, the tradeoff between the operational enhancement and the iteration number should be evaluated to assess this specific timing. For this purpose, we define a convergence-efficiency function $\eta$:

$$\eta = F(\xi) - F(T) \tag{3}$$

where $F(\xi) = \xi_k / \xi_{N_g}$, $0 \leq F(\xi) \leq 1$; $F(T) = k / N_g$, $0 \leq F(T) \leq 1$. $N_g$ represents the total number of iterations when global optimum is achieved. $k$ indicates the $k^{th}$ iteration during the GA processing, where $k \leq N_g$. $F(\xi)$ here indicates the normalized convergence as defined above. Both $F(\xi)$ and $F(T)$ are normalized to the condition when global optimum is recognized, so the range of $\eta$ would be $[-1,1]$. The convergence-efficiency function $\eta$ evaluates difference between the ratio of enhancement and the ratio of iteration number (consuming time) regarding the condition of global optimum. It presents the timing efficiency during optimization process. Hence, the maximum value of function $\eta$ could help us define the ending point of the GA processing, upon which the acquired optimized mask at this ending point should be utilized without consuming too much time.

The convergence-efficiency functions for simulations and experiments are calculated, which are shown in Fig. 5. For three different decay ratios, the convergence-efficiency functions increase from near zero to their peak values, then monotonically decrease. For numerical simulations, when decay ratio is high, the measured $\eta$ increases from the first iteration and reaches a maximum of 0.64 at the $500^{th}$ iteration, then $\eta$ reduces to about 0 again at the $2000^{th}$ iteration. For medium and low decay ratio, maximum $\eta$ appear at around $630^{th}$ iteration and $400^{th}$ iteration, the corresponding values are approximately 0.35 at the $630^{th}$ iteration, and 0.28 at the $400^{th}$ iteration, respectively. The experimental measured $\eta$ generally agrees with simulation results. For high decay ratio, maximum value locates near the $500^{th}$ iteration as around 6.0. As for medium decay ratio, the trend of $\eta$ is a little higher than simulation before the $800^{th}$ iteration, while maximum $\eta$ appears near the $630^{th}$ iteration. For the low decay ratio condition, the experiment trend is lower than simulation, with a maximum value of $\eta$ around 0.2. The results for simulations and experiments show that higher decay ratio $\Upsilon$ of mutation rate has better tradeoff relation between iteration number and enhancement of focal spot, which reflects the advantage of high decay rate from another aspect. In our assessment, the convergence-efficiency function degrades to negative after the $1500^{th}$ iteration for low decay ratio, because the enhancement increases to global optimum slowly with a large number of iterations. The negative $\eta$ indicates the terrible tradeoff between enhancement and iteration number, which should be avoided in applications. In addition, it can be aware that the iteration numbers in terms of maximum $\eta$ appear closed for high decay ratio (about $500^{th}$ iteration) and low decay ratio (about $400^{th}$ iteration), and both cases appear before the medium decay ratio (around $630^{th}$ iteration). This phenomenon might be contributed to the values of improved speeds of convergence and the improved speed change through three stages. For the low decay ratio, the convergence drastically flattens after the initial rising stage, which leads to the sudden turning point of $\eta$ at initial iterations. For the high and medium decay ratios, although the improved speeds of convergence also fall after the initial stage, they present the better approaching trends. In other words, the values of iteration number corresponding to the maximum $\eta$ may not simply proportional to the decay ratios of mutation rate. But still, the higher decay ratios could lead to larger maximum $\eta$. In some situations, the global optimum might not be accomplished, in which $N_g$ could not be determined for the evaluation of $\eta$. It is recommended to adopt the last iteration number of optimization processing as $N_g$ to evaluate $\eta$ under these situations. The convergence-efficiency function $\eta$ could be employed to analyze tradeoff relations for various optimization algorithms, rather than exclusive for genetic algorithm in this study.

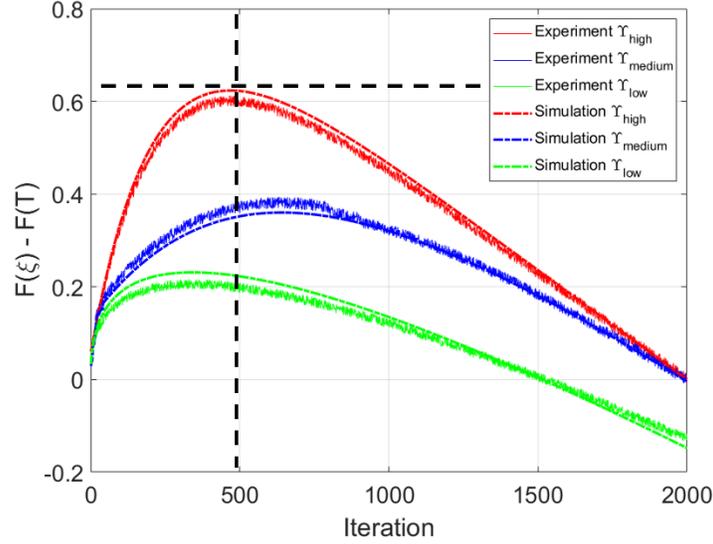

Fig. 5 Convergence-efficiency functions for simulations and experiments. Red dots, blue dots, and green dots correspond to high, medium and low decay ratios, respectively, for simulation results. Red, blue and green lines represent experimental measured convergence-efficiency functions for high, medium and low decay ratios. The black dotted line indicates the maximal point of function in experiment and simulation.

## 3. FPGA Accelerated Genetic Algorithm

In above sections, by optimizing decay ratios of mutation rate to adjust the search space, we could speed up the convergence of the GA. The simulation and experiment evaluations show that within only about 500 iterations, the normalized convergence reaches about 88% with higher decay ratio. Aiming at the applications in dynamic scattering, convergence speed is far important than the eventual enhancements regarding to the global optimum. Therefore, it is necessary to evaluate tradeoff between iteration number and operational enhancement. From our studies in both simulations and experiments, we introduce convergence-efficiency function as the criterion to judge the readiness of the GA processing as explained in the above sections. To further accelerate the GA processing for applications in dynamic scattering, minimizing the time duration for each iteration becomes another critical factor.

Based on the availabilities on DLPLCRC410EVM, we design an FPGA-based GA architecture on the Virtex-5 FPGA chip as shown in Fig. 6. The Virtex-5 FPGA from Xilinx on the board is used to control the whole program. A PLL (phase locked loop) module is generated to provide the timing control of FPGA program, including the clocks of internal state machine logic and state machine of other hardware such as DDR (Crucial PC2-6400 2GB DIMM 800MHz DDR2 SDRAM Memory) and ADC (analog-to-digital converter, LinOptx Digitizer v1.0). The top-level state machine controls the whole logic of program and cooperates with lower-level state machines. These lower-level state machines are separated as the logic of Genetic Algorithm, and data communication of DDR, ADC. In addition, operations of hardware are designed with careful considerations for the timing constraints, in which the DMD displaying interface, DDR write/read interface, and ADC read interface are designed to cooperate the procedure of the GA.

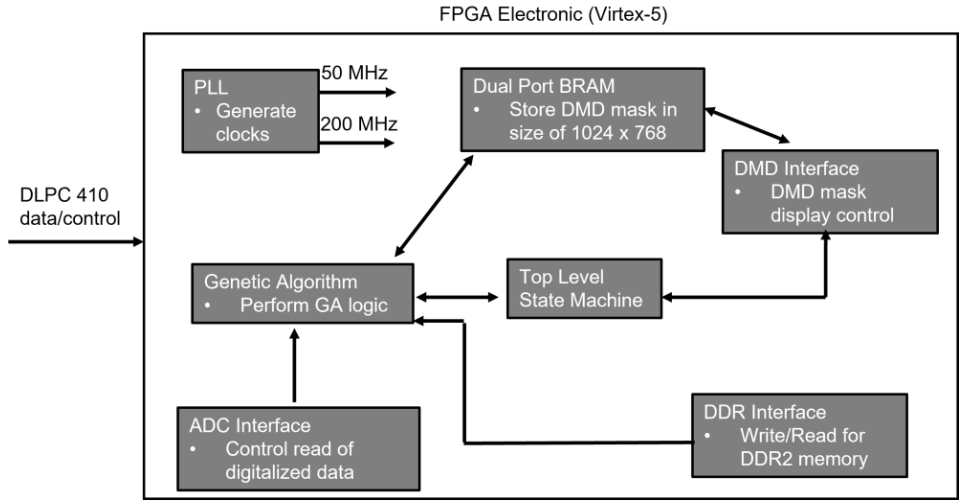

Fig. 6 Architecture in the FPGA design. PLL generates 50MHz clocks for top-level state machine, 200MHz for lower-level state machine. Dual port BRAM is an internal memory space that stores one DMD mask. DMD interface controls the displaying of masks when the full mask in size of $1024 \times 768$ is already loaded. ADC interface acquires digitalized data in high speed sampling mode. DDR interface controls the write/read operations between FPGA and DDR memory.

In the FPGA design, a trivium stream cypher is introduced to generate random number vector [24] for the randomizations. The workflow for FPGA program is shown in Fig. 7. The internal frequency for lower-level state is 200MHz with one clock cycle of 5ns. The GA starts with the generation of population of 16 parent masks, these masks are written into DDR memory for storage. The ranking process is triggered when masks are displayed on the DMD, and corresponded intensity values are captured by the photodetector, converted by the ADC, then sent into FPGA. After this initialization, the program continues to generate next population of 16 offspring masks from the high ranked parent masks that read from DDR memory. For each offspring mask, it requires one clock cycle to generate one batch of 128-bit vector, the basic size of data transferring package between FPGA and DMD buffer. Thus, the construction of one $1024 \times 768$ mask costs $8 \times 768$ clock cycles. However, since the DDR operation has 'burst' issue while reading data at slower frequency (50MHz) compared to the internal frequency of the board (200MHz), the implementations of crossovers and mutations are also only be processed based on 128-bit vectors at slower frequency. As a result, it takes $420 \mu s$ to generate one complete offspring mask, which is much longer than duration of $8 \times 768$ clock cycles. When an offspring mask is completely written to DMD buffer, the DMD starts to display the random mask immediately. During the displaying procedure of DMD, the generated offspring mask is written to DDR memory as a new parent mask for the next iteration. After that, digitalized data from ADC is accumulated in $31 \mu s$, which converts analog voltage value in range of 0~3.3V to 10-bit digital data in every $2.3 \mu s$. A $10 \mu s$ delay is set to wait for the ranking process that evaluates the fitness of offspring mask. The process for one offspring mask is finished after ranking process. For the population size of 16, the above process is conducted for 16 times as a completed iteration in GA approach, and one iteration takes 8ms.

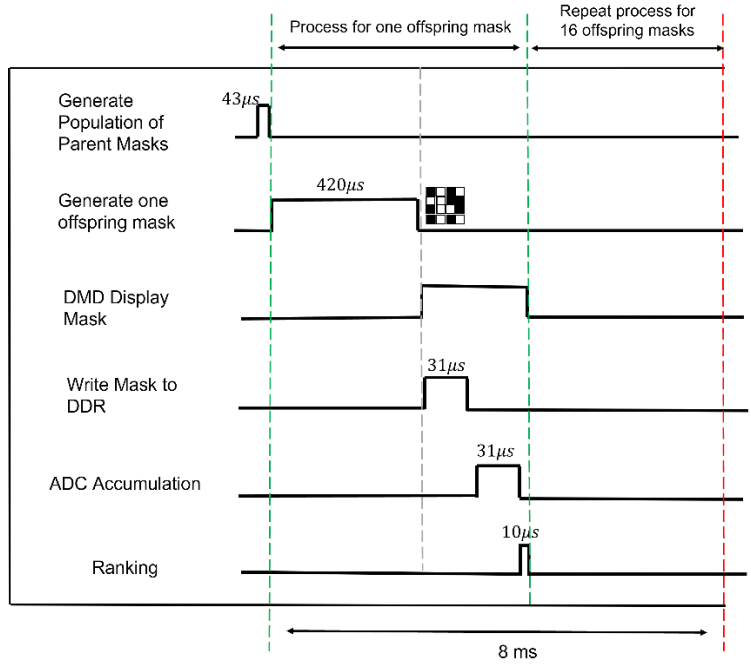

Fig.7 Workflow of generation of one offspring mask. In the beginning of program, it generates initialization parent`s population in $43\mu s$. The FPGA program starts to iterate after initialization, from where the offspring mask is constructed based on parent masks. The new offspring mask would be displayed on DMD and written to DDR, in the meanwhile, digitalized data from ADC is acquired and ranked. The process for one offspring mask cost around $500\mu s$ to complete. One iteration for processing of population of 16 is then finished in 8ms.

Considering that binary vector is more efficient to be processes in FPGA program, we modify the exponential function of mutation rate in Eq. (2) to a linear format:

$$R = \begin{cases} \left(\kappa_{start} - (k-1) \times \tau\right)/\varepsilon, & R > R_{end} \\ R_{end}, & R \leq R_{end} \end{cases} \quad (4)$$

Here $\varepsilon$ is a factor of dominator, $\kappa_{start}$ is a starting factor for the mutation rate, $\kappa_{start}/\varepsilon$ determines the initial mutation rate $R_0$. $\tau$ is a linear decay factor of the mutation rate, which is the key factor that decides the decay ratio of mutation rate. $k$ represents index of iteration. $R_{end}$ is the mutation rate at the end of the processing. To fit the high decay ratio (low $D$) acquired in the above section, the parameters are set for following values, $\kappa_{start} = 2000$, $\tau = 12$, $\varepsilon = 2^{15}$, and $R_0 = 0.061$, and $R_{end} = 0.012$.

The number of modulation units is set as $64 \times 64 = 4096$, in other words, the pixels of $1024 \times 768$ on the DMD is divided into $64 \times 64$ segments. The number of iterations is set as 2000. The experiment setup is shown in Fig. 5, where the red dotted line indicates the connection path of ADC, FPGA, and DMD. The sampled data is captured by CCD and DAQ for visualization and verification. A typical experiment result of FPGA program is presented in Fig. 8, showing the sampled voltage values from the photodetector via DAQ during the process of FPGA program. The program starts at 0s and ends at 14s. It can be seen that the improved speed of convergence experiences three stages, initial (steep rising), the second (approaching), and the third (gradually saturating), like numerical simulation results and experiment results. The green dotted line indicates the time at 4s with the 500$^{th}$ iteration, marking the end of the second stage of improved speed and corresponding to the maximal point in convergence-efficiency function shown in Fig. 5 at the 500$^{th}$ iteration. The actual enhancement is 91 at the 2000$^{th}$ iterations, indicating the global optimum of the GA. At the 500$^{th}$ iteration, the normalized convergence arrives 88%, acquired by comparing the enhancement of focal spot between the points at the 500$^{th}$ and the 2000$^{th}$ iteration. In addition, the insets in Fig. 8 display the images taken before the optimization (the speckle distribution behind scattering medium), the focal spot at the 500$^{th}$, and the focal spot at the 2000$^{th}$ iteration, respectively. This FPGA program takes a total of about 14s to finish 2000 iterations, while it only costs around 4s to reach 88% of full convergence. As contrast, the PC-based GA takes around 1200ms to complete one iteration, while our FPGA program costs 8ms. The configured FPGA program accelerates the GA processing up to 150 times.

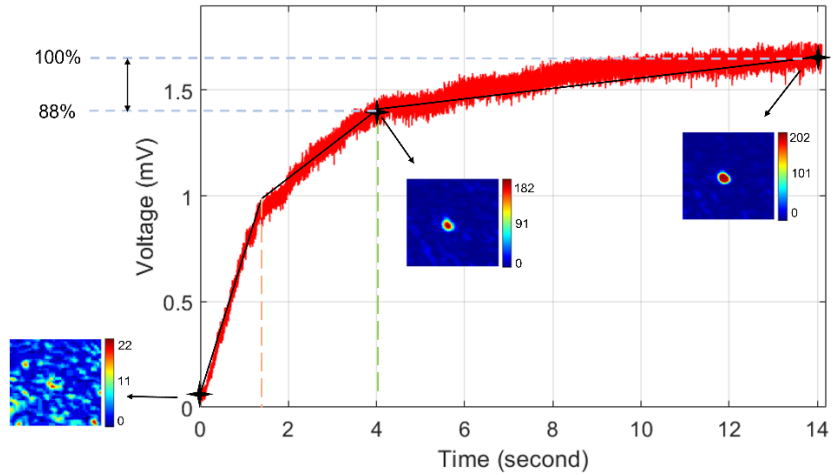

Fig. 8 Process of focusing light through scattering medium. The process of FPGA-parameter cooperative GA is recorded by CCD and DAQ. Red dots show original voltage data from DAQ. Black lines indicate three stages of convergence rate. Orange dotted line: end point of the first stage of convergence rate. Green dotted line: end point of the second stage, as the 500$^{th}$ iteration. The three inset patterns show the CCD images at the start of program, the 500$^{th}$ iteration, and 2000$^{th}$ iteration, respectively.

In order to show the possible capability for dynamic scattering medium, we modify the FPGA program to repeat GA process in every 500 iterations. Fig. 9 (a) shows results recorded via DAQ for repetition of 10 times in 40s. The insets show normalized CCD images taken at the 0s and 3.91s, corresponding to the speckle pattern without beam shaping and formed focal point at the 500$^{th}$ iteration. Fig. 9 (b) shows the cross sections of intensity distributions of the focal spots for 10 times repetition, the peak intensity value fluctuates near 200 with variance of $\pm 10$, demonstrating the robustness of our designed FPGA-based GA. For the visualization purpose, the recorded movie is presented in Supplementary Material (see Visualization 1). As a result, the FPGA program is ready for focusing light through dynamic scattering medium.

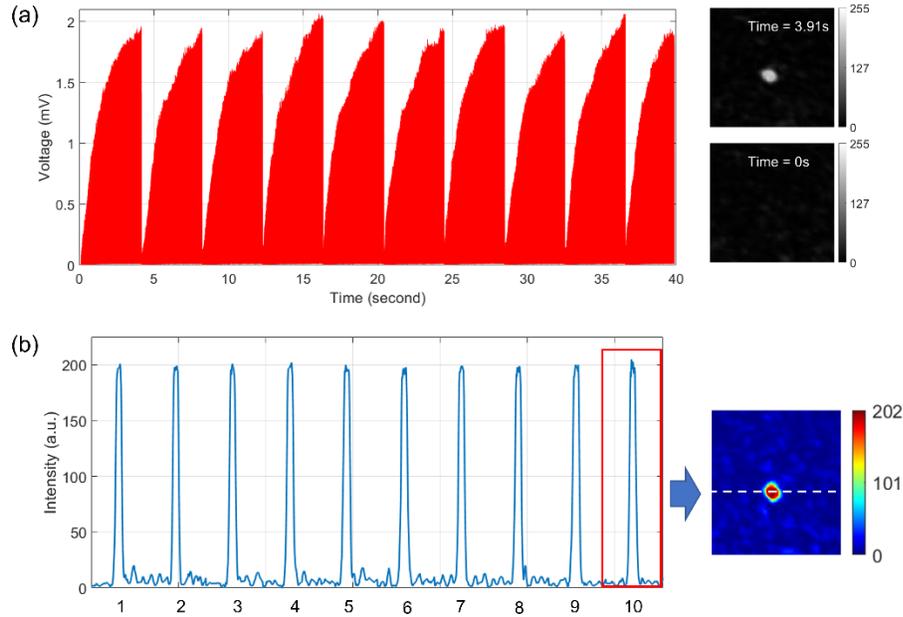

Fig. 9 Focusing light iteratively by accelerated GA. (a) Voltage data recorded by DAQ for 10 times repetition of GA. The insets figures display normalized CCD images that are captured at 0s and 3.91s. (b) The cross section of focal spot at 4s of each repeat. The inset figure shows the CCD image for the last repeat that corresponds to red rectangle area.

## 4. Discussion

Our study shows that a higher decay ratio $\Upsilon$ of mutation rate makes convergence of GA faster, but the decay ratio might not be tunned as high as possible. When decay ratio increases, the mutation rate drops quicker from the starting to the ending value. The iteration number may be insufficient to search an operable optimized mask to form a focal spot, because the mutation rate would be too low and could not provide enough search space. In our study, we choose the reasonable values of decay ratio $\Upsilon$ and initial mutation rate so that at least 100 iterations can be guaranteed before reaching the ending mutation rate. Generally, the optimal selection of $\Upsilon$ varies from case by case. It is necessary to apply MATLAB-based experimental analysis for the selection before configuring the FPGA design.

In our implementation of FPGA-parameter accelerated GA, the number of input modulation units is set as $64 \times 64$. Theoretically, the beam shaping's enhancement of the formed focal spot is proportional to the number of modulation units. However, timing limitation in the current FPGA chip in our setups, like data writing and reading speed between DDR and FPGA, prevents us from increasing the modulation units without extending the time duration for each iteration. It currently takes 14s to operate 2000 iterations for the designed FPGA based GA program. The data communication of the DDR on the board is processed at 50MHz, which is slower than the frequency of the internal state machine at 200MHz. The confliction of frequency leads to the main timing limitation of our design. The on-chip memory of Virtex-5 in the DLPLCRC410EVM from TI is only 16.4Mb. Current design is processes with population size of 16, each mask in one population is in size of $1024 \times 768$, these parameters make the design require at least memory space $1024 \times 768 \times 16 \times 2 \approx 25.1 \text{Mb}$. Consequently, the DDR must be utilized to store generated random masks. Fortunately, Xilinx now supplies FPGA chips (Virtex-UltraScale+) that enable up to 500Mb on-chip storage, which is definitely sufficient for the storage of masks for the current design. With the abundant on-chip memory, the FPGA could process all data writing/reading internally. We make a comparison between these two FPGA chips regarding the timing workflow for generating one set of 128-bit data of offspring mask (Fig. 10). The internal clock is 200MHz, meaning one clock cycle corresponds to 5ns. For the Virtex-5 FPGA and DDR hardware support in the DLPLCRC410EVM, shown in the red rectangle, the FPGA reads one set of 128-bit data from DDR with 50MHz, taking 20ns. Due to the 'burst' issue of DDR, it will take a total of 80ns to generate 128-bit offspring data. As for the Virtex-UltraScale+, shown in the dotted line rectangle, no DDR is needed to store the masks as the on-chip memory is enough. It takes 5ns to read 128-bit data from internal memory, and only 10ns in needed to generate 128-bit offspring data. If employed, the GA processing based on the Virtex-UltraScale+ FPGA chip could speed up at least eight times faster than current Virtex-5 based design. In other words, without any other improvement, it will only take 500ms or shorter to operate 500 iterations, reaching 88% convergence. Besides, the internal processing frequency of FPGA program also has potential to be improved with newer version of FPGA chip, we believe GA could be further accelerated, even with more modulation modes for higher enhancement factors. We believe that the designed GA program applied on the advanced FPGA chip would be applicable in imaging through the dynamic scattering media, which have decorrelation time on millisecond scale.

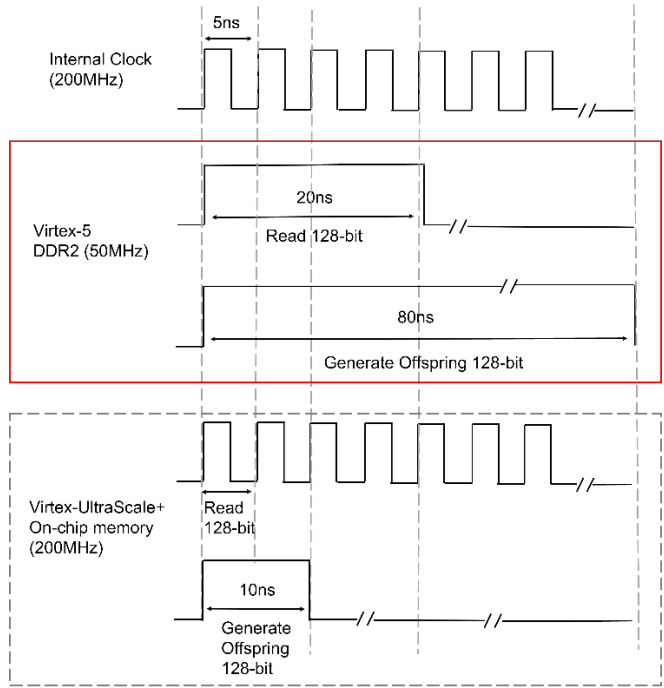

Fig. 10 Flowchart for the construction of a 128-bit data from a offspring mask by Virtex-5 and Virtex-UltraScale+. The frequency of internal clock is 200MHz with 5ns clock cycle. Red rectangle: For current Virtex-5 and DDR communication system, it takes 4 clock cycles (20ns) to read 128-bit parent data from DDR, and 80ns to generate a 128-bit offspring data. Gray rectangle: The parent data could be saved in internal memory, which reads 128-bit data out per clock cycle, thus the construction of a 128-bit offspring data requires 10ns.

## 5. Conclusion

In this study, we propose a method to achieve fast light focusing through scattering medium by Genetic Algorithm that is accelerated by optimized parameter settings and FPGA-based hardware implementation. We find that the high decay ratio improves the convergence of algorithm, which is verified in numerical simulation and experiment. By measuring the convergence-efficiency function that evaluates the tradeoff between iterations and enhancement, GA gets optimal efficiency at the 500$^{th}$ iteration, reaching 88% of the convergence. Furthermore, the FPGA program is developed to accelerate GA cooperatively at the hardware level with optimized decay ratio. The configured FPGA program implements hardware controls (DMD, ADC, DDR) and algorithm logic simultaneously. The measurement of convergence-efficiency function allows us to judge the optimization within 4s, which could be applied for dynamic scattering media of the corresponding decorrelation time of seconds. The designed GA program could be further accelerated to a millisecond scale by using the latest version of FPGA chip that provides higher on-chip memory up to 500Mb. Our study demonstrates a complicated optical system based on hardware, we believe it could pave the way for finding wide applications in biological imaging, which suffer from intense noise and short decorrelation time and thus promotes applications for real-time wavefront shaping.

**Funding**

This work was supported by the R&D funding from LinOptx. This work was partially supported by grants from The National Natural Science Foundation of China (61675140) and Science Specialty Program of Sichuan University (Grand No. 2020SCUNL210).

**Disclosures.** The authors declare no conflicts of interest.